\documentclass[twocolumn,floatfix,preprintnumbers,nofootinbib,superscriptaddress,longbibliography]{revtex4-2}

\usepackage{ulem}
\usepackage{bm}
\usepackage{times}
\usepackage{amssymb,amsbsy,amsmath,amsfonts}
\usepackage{graphicx}
\usepackage{float}
\usepackage{color}
\usepackage{morefloats}
\usepackage{rotating}
\usepackage{srcltx}
\usepackage{slashed}
\usepackage{subfigure}
\usepackage{multirow}
\usepackage{verbatim}
\usepackage{hyperref}
\usepackage{tabularx}

\usepackage[outdir=./]{epstopdf}

\usepackage{graphicx}
\usepackage{amsmath}
\usepackage{amsfonts}
\usepackage{amssymb}
\usepackage{color}
\usepackage{multirow}
\usepackage{mathrsfs}
\usepackage{gensymb}% \degree order needed

\usepackage{booktabs}  
\usepackage{threeparttable}

\usepackage{subfigure}
\newcommand\be{\begin{eqnarray}}
\newcommand\ee{\end{eqnarray}}
\newcommand{\reaction}{\pi^-p\to K^0\Lambda}
\newcommand{\gev}{\mathrm{GeV}}
\newcommand{\mev}{\mathrm{MeV}}
\newcommand{\dd}{\mathrm{d}}
\newcommand{\Mcal}{\mathcal{M}}
\newcommand{\lab}{\mathrm{lab}}
\newcommand{\cm}{\mathrm{CM}}

\usepackage{soul}
\begin{document}

\title{Reggeized \texorpdfstring{\(K^*(892)\)}{K*(892)} exchange in
\texorpdfstring{\(\pi^-p\to K^0\Lambda\)}{pi-p to K0 Lambda}
and implications for fixed-target measurements}

\author{Hao-Nan Wang}
\affiliation{State Key Laboratory of Heavy Ion Science and Technology, Institute of Modern Physics, Chinese Academy of Sciences, Lanzhou 730000, China} 
\affiliation{School of Nuclear Sciences and Technology, University of Chinese Academy of Sciences, Beijing 101408, China}
\affiliation{Few-body Systems in Physics Laboratory, RIKEN Nishina Center, Wako, Saitama 351-0198, Japan}

\author{Cheng Chen}
\affiliation{State Key Laboratory of Heavy Ion Science and Technology, Institute of Modern Physics, Chinese Academy of Sciences, Lanzhou 730000, China} 
\affiliation{School of Nuclear Sciences and Technology, University of Chinese Academy of Sciences, Beijing 101408, China}

\author{Yue Ma}
\affiliation{Few-body Systems in Physics Laboratory, RIKEN Nishina Center, Wako, Saitama 351-0198, Japan}

\author{Ju-Jun Xie}~\email{xiejujun@impcas.ac.cn}
\affiliation{State Key Laboratory of Heavy Ion Science and Technology, Institute of Modern Physics, Chinese Academy of Sciences, Lanzhou 730000, China} 
\affiliation{School of Nuclear Sciences and Technology, University of Chinese Academy of Sciences, Beijing 101408, China}
\affiliation{Southern Center for Nuclear-Science Theory (SCNT), Institute of Modern Physics, Chinese Academy of Sciences, Huizhou 516000, China}

\begin{abstract}

The two-body \(\pi^-p\to K^0\Lambda\) reaction provides a clean channel for studying strangeness production and producing tagged hyperon \(\Lambda\) samples. We describe its total cross section over the $\pi^-$ beam momentum \(2.6 \leq p_{\pi,\lab}\leq6.0~\gev/c\) using a single Reggeized \(t\)-channel \(K^*(892)\) exchange without polarization. We also examine the implications for future fixed-target measurements. An idealized event sample is used to determine the laboratory momentum and polar-angle distributions of the charged particles from \(K_S^0\to\pi^+\pi^-\) and \(\Lambda\to p\pi^-\). The \(K_S^0\) decayed two pions are strongly forward, whereas the proton and pion from the \(\Lambda\) decay extend to larger angles and lower momenta. These distributions provide a kinematic baseline for future acceptance and tracking studies.

\end{abstract}

\maketitle

%%%%%%%%%%%%%%%%%%%%%%%%%%%%%%%%%%%%%%%%%%%%%%%%%%%%%%%%%%%%%%%%%%%%%%%%%%

\section{Introduction}
\label{sec:introduction}

Hyperon--nucleon interactions are important for understanding hypernuclear structure and dense baryonic matter. Direct scattering data remain scarce because unstable hyperons cannot be transported as conventional long-lived beams~\cite{BESIII:2024geh,ParticleDataGroup:2026mpi}. Reactions that produce tagged hyperons with controlled kinematics therefore provide useful input for hadron spectroscopy and future secondary-scattering measurements. This motivation has stimulated recent proposals for high-intensity hyperon sources based on primary production reactions~\cite{Yuan:2021yks,Yuan:2026lum}.

A complementary route is provided by intense secondary meson beams. Separated pion and kaon beam lines have been operated at facilities such as J-PARC, while a secondary neutral-kaon beam has been proposed at Jefferson Lab for strange-hadron spectroscopy~\cite{Agari:2012kid,KLF:2020gai}. At the high-intensity heavy-ion accelerator facility (HIAF), currently under construction in Huizhou, Guangdong Province, China, is expected to be completed by 2025. And, the construction of a dedicated experimental apparatus at HIAF—the Huizhou Hadron Spectrometer (HHaS) was proposed~\cite{Chen:2025ppt} for high-rate hadron measurements and its future program considers secondary pion and kaon beams with expected intensities of order \(10^{8}~\mathrm{s}^{-1}\) and \(10^{6}~\mathrm{s}^{-1}\), respectively. Such beams would provide high-statistics data for exclusive reactions with strange final states, including differential cross sections and polarization observables.

From the physics point of view, secondary pion and kaon beams provide direct probes of meson--baryon interactions~\cite{Ji:2026rin}, baryon and hyperon spectroscopy~\cite{Liu:2012ge,Cao:2014mea,Garzon:2015zva,Xie:2015zga,Lu:2015fva,Zhou:2019eaf}, and coupled-channel dynamics~\cite{LiQuZhi:2021nrq,Wang:2023snv,Ronchen:2012eg}. For example, experimental measurements of reactions such as \(\pi N\to K\Lambda\), \(K\Sigma\), \(\omega N\), and $\phi N$ constrain the corresponding partial-wave amplitudes and reduce the model dependence in the extraction of excited \(N^*\) and \(\Delta^*\) properties~\cite{Matsuyama:2006rp,Xie:2007qt,Ronchen:2012eg,Wu:2014yca,Kamano:2013iva,Wang:2022osj}. Within this broader program, the reaction \(\pi^-p\to K^0\Lambda\) is a simple exclusive channel for strangeness production. Its two-body kinematics allow direct tests of the production amplitude and provide a controlled source of tagged \(\Lambda\) hyperons. At multi-GeV pion momenta, its forward-angle behavior also provides a useful test of strange-meson exchange mechanisms, which are the focus of the present work.

The exclusive two-body reaction  \(\pi^- p \to K^0 \Lambda \) provides a particularly transparent production channel. At fixed incident
momentum, the final-state four-momenta are specified by one center-of-mass (c.m.) scattering angle and one azimuthal angle. Historical measurements provide total cross sections, angular distributions, and \(\Lambda\) polarization over a broad energy range~\cite{Dahl:1967hix,Crennell:1972km,Beusch:1974rr,Snyder:1979ya}. More recently, the J-PARC E40 experiment reported finely binned forward-angle polarization data at \(p_{\pi,\lab}=1.33~\gev/c\)~\cite{J-PARCE40:2025hhw}.

Near threshold, several \(N^*\) amplitudes can contribute. The reaction has been investigated using phase-shift and partial-wave analyses~\cite{Wagner:1971aa,Baker:1978qm,Saxon:1979xu,Anisovich:2011fc}, as well as effective-Lagrangian and dynamical coupled-channel approaches~\cite{Shklyar:2005xg,Julia-Diaz:2006rvy,Ronchen:2012eg,Kamano:2013iva,Wu:2014yca}. At higher beam momenta and small momentum transfer, forward production can be described economically in terms of \(t\)-channel exchange. The vector meson \(K^*(892)\)~\footnote{Hereafter, \(K^*\) denotes \(K^*(892)\).} connects the \(K\pi\) and \(N\Lambda\) vertices and naturally produces a forward enhancement. Replacing the ordinary Feynman pole by a Regge trajectory introduces the characteristic energy and momentum-transfer dependence of high-energy exchange \cite{Nam:2010au,Wang:2014jxb,Godizov:2015hbm,Giordano:2011sn,Basar:2017ocn,Toki:2007ab,Titov:2005kf}. The present work tests the restricted hypothesis that Reggeized \(K^*\) exchange provides the leading contribution over the selected relatively high-momentum range. After fixing the two vertex couplings, we determine whether one form-factor cutoff can describe the total-cross-section data. We also compare the Reggeized result with the same exchange amplitude evaluated using an ordinary Feynman pole propagator.

Regarding the coupling of \(K^*\Lambda N\), only the vector interaction is retained. With real couplings and no relative phase between independent production amplitudes, the model predicts zero normal \(\Lambda\) polarization, as studied for the case of $p \bar{p} \to \Lambda \bar{\Lambda}$ reaction in Ref.~\cite{Wang:2025fnf}. Polarization data are therefore not included in this work. A measured nonzero polarization would indicate that the restricted amplitude is incomplete and that additional spin structures, rescattering effects, or production mechanisms are required.

The remainder of this article is organized as follows. Section~\ref{sec:theory} presents the \(K^*\)-exchange amplitude, compares the Reggeized and ordinary propagators, and determines the form-factor cutoff from the total cross section. Section~\ref{sec:experiment} estimates fixed-target yields and identifies differential observables that can test the assumed exchange mechanism. The fitted angular distribution is then propagated through the visible \(K_S^0\) and \(\Lambda\) decay channels. This generator-level study provides a kinematic baseline for future acceptance calculations, but it does not constitute a detector simulation. Section~\ref{sec:summary}
is a short summary.

\section{Theoretical formalism and results}  \label{sec:theory}

\begin{figure}[htbp]
  \centering
  \includegraphics[width=1\linewidth]{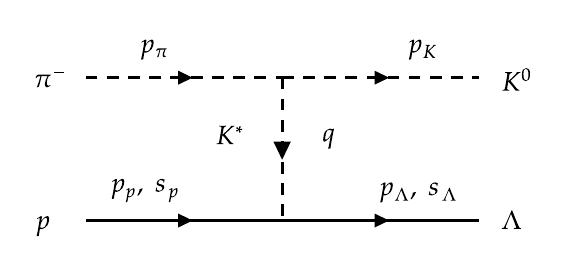}
  \caption{The $t$-channel $K^*$-exchange diagram for $\pi^- p \to K^0 \Lambda$ reaction. The definition of the kinematical ($p_{\pi}$, $p_p$, $p_{K}$, $p_{\Lambda}$) and polarization variables ($s_p$, $s_\Lambda$) that we use in the present calculation are also shown. In addition, $q = p_{\pi} - p_{K}$ denotes the four-momentum of the exchanged $K^*$ meson.}
  \label{fig:Kstar-exchange}
\end{figure}

We follow the work of Ref.~\cite{Wu:2014yca}, investigating the $\pi^- p \to K^0 \Lambda$ reaction with an effective Lagrangian approach. However, in the present work, we focus on the high energy region of that reaction and the $t$-channel $K^*$ exchange, which is depicted in Fig.~\ref{fig:Kstar-exchange}.

\subsection{Reggeized \texorpdfstring{\(K^*\)}{K*} exchange formalism}  \label{subsec:formalism}

The effective interaction Lagrangian for the $K^*K\pi$ vertex is taken as used in previous works~\cite{Xie:2010yk,Xie:2013wfa,Xie:2013db}:
\begin{equation}
  \mathcal{L}_{K^*K\pi}
  =-g_{K^*K\pi}
  \left(
  \partial^\mu \bar K \bm \tau \cdot \bm \pi
  -\bar K \bm \tau \cdot \partial^\mu \bm \pi
  \right)K^*_{\mu}.
  \label{eq:lkkpi}
\end{equation}
Within SU(3) flavor symmetry \(g_{K^* K\pi}=-g_{\rm PPV}\), one can get \(g_{K^*K\pi}=-3.02\)~\cite{Wu:2014yca,Ronchen:2012eg}. Then, in the convention
\begin{equation}
  \bm\tau\cdot\bm\pi=
  \begin{pmatrix}
    \pi^0 & \sqrt{2}\pi^+\\
    \sqrt{2}\pi^- & -\pi^0
  \end{pmatrix},
\end{equation}
the charged vertex relevant to \(\reaction\) carries a factor \(\sqrt{2}\).
We therefore use
\begin{equation}
  g_{K^*K\pi}^{\mathrm{charge}}
  =\sqrt{2}\,g_{K^*K\pi}
  =-4.27.
  \label{eq:charge-meson-coupling}
\end{equation}
Because this charge-channel coupling is inserted directly into the amplitude, no additional isospin factor is applied.

For the $K^*N\Lambda$ vertex, we use the following effective interaction Lagrangian which only the vector interaction is included:
\begin{equation}
  \mathcal{L}_{K^*\Lambda N}
  =g_{K^*\Lambda N}\bar N \gamma^\mu K^*_{\mu} \Lambda + \mathrm{H.c.}
  \label{eq:lbaryon}
\end{equation}
The SU(3) flavor symmetry gives $g_{K^*\Lambda N} =-6.19$, which is following the convention of Ref.~\cite{Wu:2014yca}.

With the effective interaction Lagrangian densities given above, we can easily construct the invariant scatttering amplitude,
\be
  \Mcal_{s_\Lambda s_p}(s,t)
   && = g_{K^*K\pi}^{\mathrm{charge}}g_{K^*\Lambda N}
  \bar u_\Lambda(p_{\Lambda},s_\Lambda)\gamma^\mu u_p(p_p,s_p)
  \nonumber\\
  && \!\!\!\!\!\!\!\!\!\!\!\! \times\left(-g_{\mu\nu}
  +\frac{q_\mu q_\nu}{m_{K^*}^2}\right)
  (p_{\pi}+p_{K})^\nu
  \mathcal{P}_{K^*}(s,t),
  \label{eq:amplitude}
\ee
where $t = q^2$, and \( s=(p_{\pi}+p_p)^2 \) is the squared center-of-mass energy. For the main calculation at high energies, the ordinary \(K^*\) pole propagator is replaced by a Regge trajectory propagator~\cite{Navelet:1976ht,Guidal:1997hy,Guidal:1997by,Laget:2005be},which takes into account the exchange of a {\it family} of $K^*$ mesons, and we take a linear trajectory for the exchanged $K^*$ meson as used in previous works~\cite{Brisudova:1999ut,Titov:2005kf,Toki:2007ab,Nam:2010au}:
\begin{equation}
  \alpha_{K^*}(t)
  =1+\alpha'_{K^*}(t-m_{K^*}^2),
  \qquad
  \alpha'_{K^*}=0.9~\gev^{-2},
  \label{eq:trajectory}
\end{equation}
and the Reggeized propagator
\be
  \!\!\!\!\! \mathcal{P}_{K^*}^{\rm Regge}(s,t)
  =\left(\frac{s}{s_0}\right)^{\alpha_{K^*}(t)-1}
  \frac{\pi\alpha'_{K^*}}
  {\Gamma[\alpha_{K^*}(t)]
   \sin[\pi\alpha_{K^*}(t)]},
  \label{eq:regge}
\ee
where \(s_0=1~\gev^2 \) is a mass scale, and we have followed the convention of Refs.~\cite{Nam:2010au,Titov:2005kf,Toki:2007ab}, where a constant phase equal to unity is adopted, and no rotating signature phase is included.

For comparison, we also evaluate the scattering amplitude of $\pi^- p \to K^0 \Lambda$ reaction with the ordinary Feynman pole propagator, which is
\begin{equation}
  \mathcal{P}_{K^*}^{\rm Feynman}(t)
  =\frac{1}{m_{K^*}^2-t}.
  \label{eq:feynman-propagator}
\end{equation}
Note that the width of $K^*$ is not included in Eq.~\eqref{eq:feynman-propagator}. In the physical region considered here, the exchanged \(K^*\) is a \(t\)-channel virtual particle with $t<0$ and does not approach its time-like mass shell. Equation~\eqref{eq:feynman-propagator} is therefore an ordinary pole propagator rather than a resonant Breit--Wigner form~\cite{Wang:2025fnf}. For the $K^*$ mass, we use $m_{K^*(892)} =(m_{K^{*+}}+m_{K^{*0}})/2 = 893.6~\mev$, where the hadroproduced mass of $K^{*+}$ meson is adopted, as given by the Review of the Particle Physics (RPP)~\cite{ParticleDataGroup:2026mpi}.

To parametrize the off-shell behavior of the exchanged \(K^*\) meson, we multiply the scattering amplitude by a monopole form factor
\begin{equation}
  F(t,\Lambda_{\rm cut})
  =\frac{\Lambda_{\rm cut}^2-m_{K^*}^2}
  {\Lambda_{\rm cut}^2-t},  \label{eq:f-amplitude}
\end{equation}
as commonly used in effective Lagrangian models~\cite{Xie:2014twa,Xie:2015zga,Lu:2015fva}. It is worthy to mention that the cutoff \(\Lambda_{\rm cut}\) is the only fitted parameter in the present work.

For an unpolarized proton target and spinless initial $\pi^-$ and external $K^0$ mesons, the spin-averaged squared scattering amplitude for the $\pi^- p \to K^0 \Lambda$ reaction is written as,
\begin{equation}
  \overline{|\Mcal|^2}
  = F^2(t,\Lambda_{\rm cut}) \frac{1}{2} \sum_{s_p=\pm1/2}
  \sum_{s_\Lambda=\pm1/2}
  \left|\Mcal_{s_\Lambda s_p}\right|^2.
  \label{eq:spin-average}
\end{equation}
The factor \(1/2\) averages over the two initial-proton spin states, while the final \(\Lambda\) spin is summed. 

Then, the differential cross section is written as~\cite{ParticleDataGroup:2026mpi},
\begin{align}
  \frac{\dd\sigma}{\dd \cos\theta_K^{\cm} }
  &=\frac{|\boldsymbol{p}_{K}|}{32\pi s |\boldsymbol{p}_{\pi}|}
  \overline{|\Mcal|^2},
  \label{eq:dsigmadz}
\end{align}
where \(\theta_K^{\cm}\) is the angle between the incident pion and outgoing $K^0$ meson in the c.m. frame. And the three-momentum magnitudes of initial $\pi^-$ and final $K^0$ mesons are
\begin{equation}
  |\boldsymbol{p}_{\pi}|
  =\frac{\lambda^{1/2}(s,m_\pi^2,m_p^2)}{2\sqrt{s}}, ~
  |\boldsymbol{p}_{K}|
  =\frac{\lambda^{1/2}(s,m_K^2,m_\Lambda^2)}{2\sqrt{s}},
  \label{eq:cm-momentum-magnitudes}
\end{equation}
where
\begin{equation}
  \lambda(x,y,z)=x^2+y^2+z^2-2xy-2xz-2yz
  \label{eq:kallen-function}
\end{equation}
is the K\"all\'en function.

Regarding the polarization of the \(\Lambda\) hyperon,  we can inspect the spin structure from spinor definitions, which can be given by
\begin{equation}
u_i(p_i, \lambda_i) = \frac{1}{\sqrt{E_i+m_i}} (\slashed{p}_i+m_i) \frac{1}{2} 
  (1 + \lambda_i \gamma^5 \slashed{s}_i) {u_i}_0,
\end{equation}
where \( {s}_i \) is the spin four-vector satisfying $s_i \cdot s_i=-1, s_i \cdot p_i = 0$, $\slashed{s}_i\equiv\gamma^\mu s_{i\mu}$, and \(\lambda_i = \pm 1\) denotes the helicity. Here \({u_i}_0\) is an arbitrary reference spinor, as long as it satisfies the normalization condition for \( u_i(p_i, \lambda_i) \). The subscript $i$ labels the particle species. Thus, we have
\begin{align}
  & \frac{\slashed{p}_i+m_i}{2m_i} \frac{1 + \lambda_i \gamma^5 \slashed{s}_i}{2}
  u_i(p_i, \lambda_i) = u_i(p_i, \lambda_i), \label{eq:spin}\\
  & \frac{1 + \alpha \gamma^5 \slashed{s}_i}{2} \frac{1 + \beta \gamma^5 \slashed{s}_i}{2}
  = \delta_{\alpha\beta} \frac{1 + \alpha \gamma^5 \slashed{s}_i}{2}. \label{eq:ort}
\end{align}
For fixed initial and final spins, the transition amplitude is given by \(T_{\alpha, \lambda} = \bar{u}_i(p_i, \alpha) \slashed{p}_i^{\mu} u_i(q_i, \lambda)\) and the spin density matrix \(\rho_{\alpha\beta}\) is proportional to \(\sum_{\lambda} T_{\alpha, \lambda} T^*_{\beta, \lambda}\). Using Eqs.~\eqref{eq:spin} and \eqref{eq:ort}, one can show that the spin density matrix is proportional to the identity matrix. This implies that the normal polarization of $\Lambda$ vanishes, i.e.,
\begin{equation}
  P_y = \frac{\dd \sigma_{s_{\Lambda}=+n} - \dd \sigma_{s_{\Lambda}=-n}}
  {\dd \sigma_{s_{\Lambda}=+n} + \dd \sigma_{s_{\Lambda}=-n}} = 0.
\end{equation}
This is a boundary of the restricted model, not a statement that every
high-energy \(\Lambda\) sample must be unpolarized. Taking into account
the qualitative estimation of the total cross section and limited impact on 
the kinematics of decay products from the polarization,
it's reasonable to neglect the polarization in the present study.

\subsection{Numerical fitting results} \label{subsec:fitting_results}

We have performed a $\chi^2$ fit to experimental measurements for the total cross sections of the $\pi^- p \to K^0 \Lambda$ reaction at pion laboratory
momenta from \(2.6\) to \(6.0~\gev/c\)~\cite{Dahl:1967hix,Miller:1965Strange,Armenteros:1964zza,Goussu:1967Strange,Abramovich:1971pv,CERN-CollegedeFrance-Madrid-Stockholm:1980nhd,Crennell:1972km}. There are 13 experimental data points. The fitted parameter and the corresponding‌ $\chi^2/N_{\rm dof}$ are as following:
\begin{equation}
  \Lambda_{\rm cut}=1381.2 \pm 30.7~\mev,~~~~
  \chi^2/N_{\rm dof}=1.0,
  \label{eq:fit-result_reggeized}
\end{equation}
And, the fitting result with the ordinary $K^*$ Feynman propagator is
\begin{equation}
  \Lambda_{\rm cut}'=928.2 \pm 1.4~\mev,~~~~
  \chi^{'2}/N_{\mathrm{dof}}=18.6.
  \label{eq:fit-result_feynman}
\end{equation}
We find the first fit has a reasonably small $\chi^2$, while the second fit has a relative larger $\chi^2$. This poor fitting quality shows that the ordinary \(K^*\) propagator as in Eq.~\eqref{eq:feynman-propagator} cannot describe the total cross section over the fitted range
within the current model. The value of \(\chi^2/N_{\mathrm{dof}}\) shows that the restricted Reggeized \(K^*\)-exchange model can reproduce the total-cross-section data of the $\pi^- p \to K^0 \Lambda$ reaction. The energy dependent of the total cross sections in the range of 2.5 to 6 GeV is suggestive of Regge-type behavior for the $t$-channel $K^*$ exchange, and the monopole form factor~\footnote{In the fit, we also examined other choices for the form factor and found that the monopole type gives the smallest $\chi^2$.} of Eq.~\eqref{eq:f-amplitude}.

\begin{figure}[htbp]
  \centering
  \includegraphics[scale=0.42]{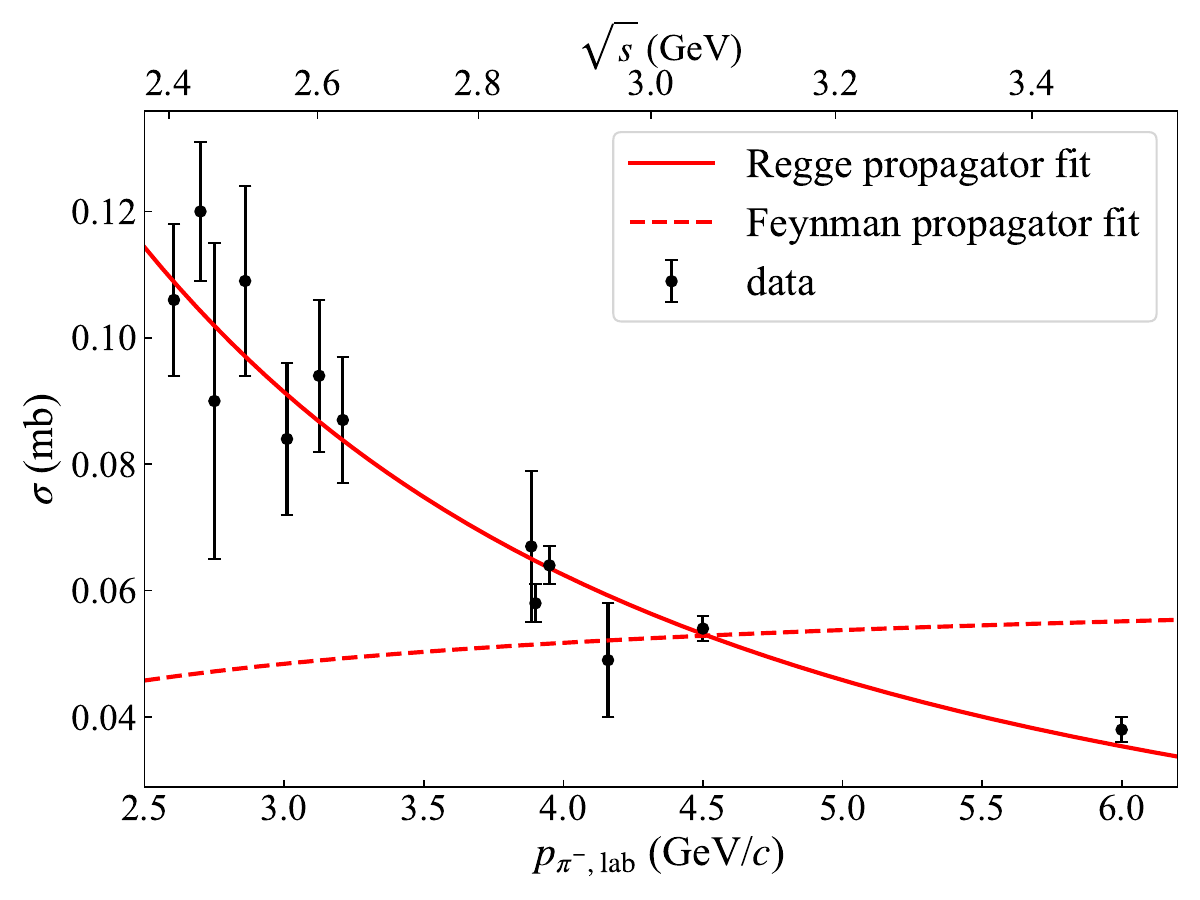}
  \caption{Total cross section for \(\reaction\). The experimental data are taken from Refs.~\cite{Dahl:1967hix,Miller:1965Strange,Armenteros:1964zza,Goussu:1967Strange,Abramovich:1971pv,CERN-CollegedeFrance-Madrid-Stockholm:1980nhd,Crennell:1972km}. The solid curve is the Reggeized \(K^*\)-exchange result obtained by fitting \(\Lambda_{\rm cut}\), while the dashed curve is the fitting results using Feynman propagator for $K^*$-exchange. Note that the invariant mass scale of the $\pi^- p$ system ($\sqrt{s}$) is also shown in the upper side of the figure.}
  \label{fig:total-fit}
\end{figure}

In Fig.~\ref{fig:total-fit}, we show the fitted total cross sections as a function of $p_{\pi,\lab}$, together with the experimental data. The calculated cross sections decreases smoothly from approximately \(0.12~\mathrm{mb}\) near the lower end of the fitted range to \(0.04~\mathrm{mb}\) at \(p_{\pi,\lab}=6~\gev/c\). This behavior is reasonably well reproduced by the dominant $t$-channel $K^*$ exchange contribution within the Regge approach, which is consistent with the analysis of the $\pi^- p \to K^0 \Lambda$ reaction from 1.5 to 4.2 GeV reported in Ref.~\cite{Dahl:1967hix}.

It is worth noting that $\alpha'_{K^*} = 0.9 ~{\rm GeV}^{-2}$ used in Eq.~\eqref{eq:trajectory} is also in consistent with the values of 0.85 and 0.8 ${\rm GeV}^{-2}$ used in Ref.~\cite{Nam:2010au} and Ref.~\cite{Toki:2007ab}, respectively. Furthermore, it is found that the total cross section exhibits an increasingly gradual decrease over the selected range as the $\alpha'_{K^*}$ decreases. This agreement does not establish that \(K^*\) exchange is the unique production mechanism, since the angular integration removes information about the scattering amplitude, and contributions omitted from the model may be absorbed partly into the fitted cutoff.

\begin{figure}[htbp]
  \centering
  \includegraphics[scale=0.4]{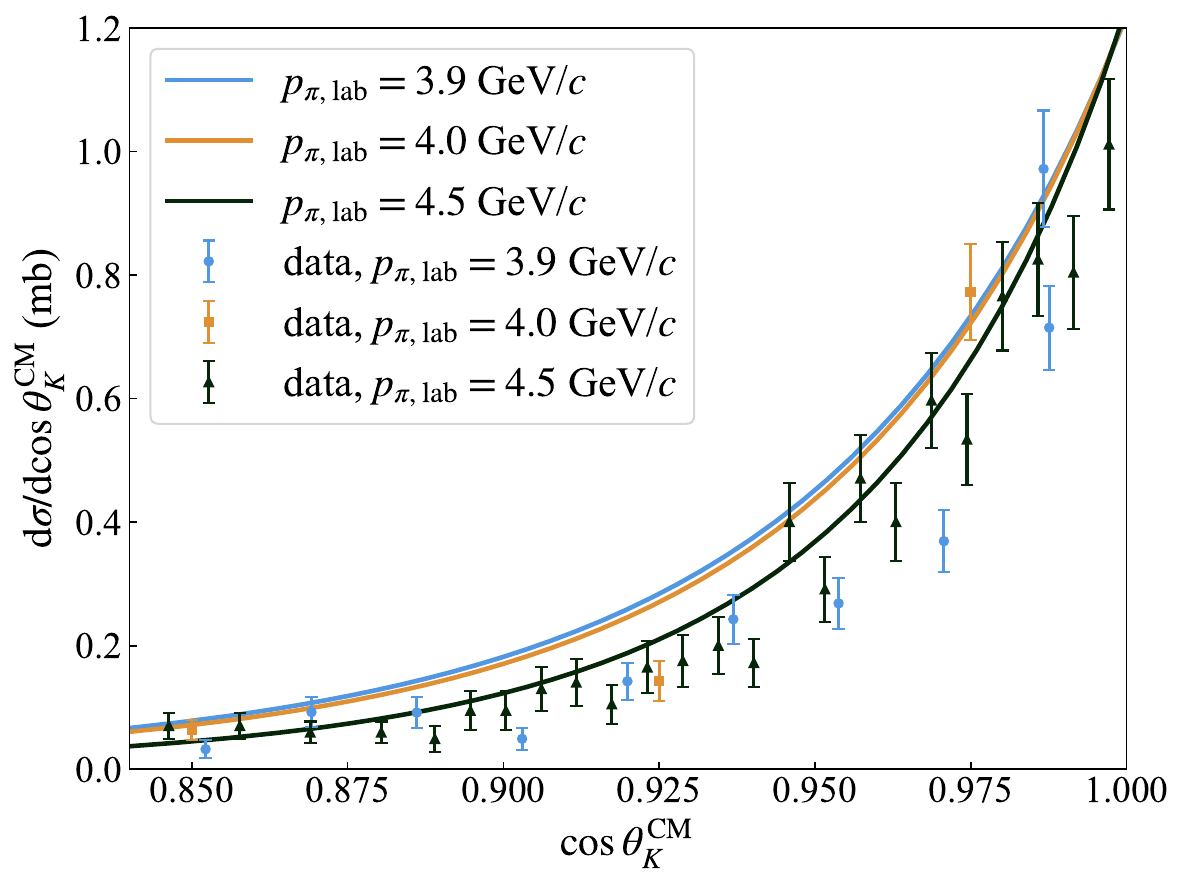}
  \caption{The differential cross section \(\dd\sigma/\dd \cos{\theta_K^{\cm}} \) calculated at 3
  representative beam momenta (3.9, 4.0, 4.5 GeV/$c$), with \(\Lambda_{\rm cut}\) from the total cross section, are compared with the corresponding experimental data~\cite{Abramovich:1971pv,Dahl:1967hix,Crennell:1972km}. These curves are not fits to differential data.}
  \label{fig:dsigma}
\end{figure}

In order to better test the present model, it is very useful to calculate the differential cross section \(\dd\sigma/\dd \cos{\theta_K^{\cm}}\) as defined in Eq.~\eqref{eq:dsigmadz}. We then, in Fig.~\ref{fig:dsigma}, compare our theoretical calculations for the differential cross sections of $\pi^- p \to K^0 \Lambda$ reaction as a function of $\cos{\theta_K^{\cm}}$ with the currently available data, at $p_{\pi, \rm lab}=3.9,~4.0,~4.5~\gev/c$. The related experimental data points from Refs.~\cite{Abramovich:1971pv,Dahl:1967hix,Crennell:1972km} are plotted for comparison, the curves are found to qualitatively follow the trend of the data and show no significant discrepancy in magnitude. The differential cross sections exhibit a pronounced enhancement in the forward-angle region, \(\cos\theta_K^{\cm}\to 1\), as expected from \(t\)-channel $K^*$ exchange at small \(-t\). At a fixed value of \( \cos{\theta_K^{\cm}} \), the differential cross section in Fig.~\ref{fig:dsigma} decreases as the incident pion momentum increases. This behavior is consistent with the overall decreasing trend of the total cross section with increasing \(p_{\pi,\rm lab}\). To better display this forward enhancement, Fig.~\ref{fig:dsigma} shows only the small forward angle region~\footnote{Note that no angular distribution data were used to constrain the model parameters.}.

\section{Experimental method and simulation}  \label{sec:experiment}

With the model parameters fixed and \(\Lambda_{\rm cut}\) determined from the total cross-section fit in Eq.~\eqref{eq:fit-result_reggeized}, the model predicts the differential cross section throughout the fitted momentum range without further adjustment. The higher statistic and more accurate measurements of \(\dd\sigma/\dd\cos\theta_K^{\cm}\) at several beam momenta would therefore provide a direct test of the assumed dominance of reggeized \(K^*\) exchange. In particular, the same parameter set should describe both
the absolute normalization and the angular shape.  

Based on the current mechanism, the differential measurements could test two complementary features of the model. The forward variation of \(\dd\sigma/\dd\cos\theta_K^{\cm}\) probes the momentum-transfer dependence of the exchange amplitude, including the vertex form factor. On the other hand, the measurements at fixed \(t\) could test the energy dependence predicted by Eq.~\eqref{eq:dsigmadz}. Together, these comparisons would help assess whether the single reggeized \(K^*\) exchange provides an adequate description for the two-body \(\pi^-p\to K^0\Lambda\) reaction with high incident pion momentum.

The normal \(\Lambda\) polarization provides a further, independent test. Although polarization data are not included in the present fit,
the model predicts \(P_y=0\). A significant nonzero measurement would indicate relative phases between independent spin amplitudes that are
absent from the present description. To connect these tests with future fixed-target measurements, we estimate the production yield and simulate the laboratory momentum and angular distributions of the charged particles from \(K_S^0\to\pi^+\pi^-\) and \(\Lambda\to p\pi^-\) decays. These distributions provide a kinematic baseline for subsequent acceptance and tracking studies.

\subsection{Production-yield estimate and statistical criterion}

We estimate the production yield of associated $K^0$ and $\Lambda$ directly from the calculated exclusive total cross section. For a liquid-deuterium or liquid-hydrogen target, denoted by \(X=\mathrm{D}_2\) or \(\mathrm{H}_2\), the production probability per incident pion in the thin-target approximation is
\begin{equation}
  P_{\mathrm{prod}}^{(X)}
  \simeq
  \sigma_{\pi^-p\to K^0\Lambda}n_p^{(X)},
  ~~~~
  n_p^{(X)}=\frac{\rho_XL_X}{M_X}N_A,
  \label{eq:production-probability}
\end{equation}
where \(\sigma_{\pi^-p\to K^0\Lambda}\) is the elementary \(\pi^-p\to K^0\Lambda\) total cross section at the incident pion momentum under consideration, and \(n_p^{(X)}\) is the areal density of target protons. Here, \(\rho_X\) and \(L_X\) denote the mass density and effective thickness of target \(X\), respectively. \(M_X\) is the molar mass defined per hydrogen or deuterium nucleus, and \(N_A\) is the Avogadro constant.

For the liquid-deuterium estimate, we use the original 10-cm target. Its density is \(\rho_D=0.169~\mathrm{g/cm^3}\), and the molar mass per deuterium nucleus is \(M_D=2.014~\mathrm{g/mol}\). For comparison, a liquid-hydrogen target of the same length is evaluated with \(\rho_H=0.0708~\mathrm{g/cm^3}\) and \(M_H=1.0079~\mathrm{g/mol}\)~\cite{ParticleDataGroup:2026mpi}. The corresponding proton areal densities are
\begin{equation}
  n_p^{(D)}=5.05\times10^{23}~\mathrm{cm^{-2}},~
  n_p^{(H)}=4.23\times10^{23}~\mathrm{cm^{-2}}.
  \label{eq:target-number}
\end{equation}

At \(p_{\pi,\lab}=4~\gev/c\), the fitted total cross section is approximately \(0.62\times10^{-28}~\mathrm{cm^2}\). Then, with Eq.~\eqref{eq:production-probability} we get
\begin{equation}
  P_{\mathrm{prod}}^{(D)}=3.13\times10^{-5},\qquad
  P_{\mathrm{prod}}^{(H)}=2.62\times10^{-5}.
  \label{eq:target-production-probabilities}
\end{equation}
The production rate and
the number accumulated over an effective beam-on time \(T_{\rm run}\) are
\begin{equation}
  R_{\mathrm{prod}}^{(X)}=\Phi_\pi P_{\mathrm{prod}}^{(X)},\qquad
  N_{\mathrm{prod}}^{(X)}=R_{\mathrm{prod}}^{(X)}T_{\rm run}.
  \label{eq:produced-yield}
\end{equation}

As an example, the HHaS design study quotes a secondary-pion intensity of order \(\Phi_\pi=10^8~\mathrm{s}^{-1}\) as shwon in Ref.~\cite{Chen:2025ppt}, thus we can use it to clarify the estimation clearly. Numerically, the raw rates are \(3.13 \times 10^3~\mathrm{s}^{-1}\) for liquid deuterium and \(2.62\times10^3~\mathrm{s}^{-1}\) for liquid hydrogen before decay and detector losses.

Because the detectors can only measure the final charged particles, it is necessary to take the transfer factors after $K^0$ and $\Lambda$ production into account. The flavor state of neutral kaon is a coherent superposition of the short- and long-lived propagation eigenstates. If neglecting the small CP-violating terms, we can use \(\lvert K^0\rangle\simeq
(\lvert K_S^0\rangle+\lvert K_L^0\rangle)/\sqrt{2}\). For a near-target \(\pi^+\pi^-\) sample dominated by the short-lived component, the fiducial probability can be written schematically as
\begin{equation}
  \mathcal P^{\rm fid}_{K^0\to\pi^+\pi^-}
  \simeq \frac{1}{2}\,
  \mathcal B(K_S^0\to\pi^+\pi^-).
  \label{eq:kaon-fiducial-probability}
\end{equation}
With the decay branching ratio of $\Lambda \to p \pi^-$, the reconstructed yield is
\begin{align}
  N_{\mathrm{rec}}={}&N_{\mathrm{prod}}
  \mathcal P^{\rm fid}_{K^0\to\pi^+\pi^-}
  \mathcal B(\Lambda\to p\pi^-)
  \epsilon,
  \label{eq:reconstructed-yield}
\end{align}
where efficiency \(\epsilon\) mainly includes the geometrical acceptance, tracking efficiency, event selection efficiency and so on. Using the current values \(\mathcal B(K_S^0\to\pi^+\pi^-)=0.692\) and \(\mathcal B(\Lambda\to p\pi^-)=0.641\) from RPP~\cite{ParticleDataGroup:2026mpi}, the ideal visible fraction before detector losses is
\begin{equation}
  \mathcal B_{\rm vis}^{(0)}
  =\mathcal P^{\rm fid}_{K^0\to\pi^+\pi^-}
  \mathcal B(\Lambda\to p\pi^-)=0.22.
  \label{eq:visible-branching-fraction}
\end{equation}
Taking this fraction parameter into account, the production rates are approximately \(6.89\times10^2~\mathrm{s}^{-1}\) for liquid deuterium and \(5.76\times10^2~\mathrm{s}^{-1}\) for liquid hydrogen. Accordingly, one day of effective beam time would produce approximately \(5.95\times10^7\) and \(4.99\times10^7\) events for liquid deuterium and liquid hydrogen, respectively.

\subsection{Event generation}
\label{subsec:3B}

According to the calculation results from Sec.~\ref{sec:theory}, we would like to illustrate a simulation of $\pi^- p \to K^0 \Lambda$ generation within the reggeized $K^*$ exchange formalism. Given the one day production amount estimated before, an idealized generator-level sample of \(5\times10^7\) unweighted events was produced at \(p_{\pi,\lab}=4~\gev/c\), with the target proton at rest frame. A trial value
\(z=\cos\theta_K^{\cm}\) was drawn uniformly from \([-1,1]\), and the event was accepted with probability
\begin{equation}
  P_{\rm acc}(z)=
  \frac{[\dd\sigma/\dd z](z)}{[\dd\sigma/\dd z]_{\rm max}}.
  \label{eq:acceptance-rejection}
\end{equation}
This is an unweighting method. The production azimuth was sampled uniformly over \([0,2\pi)\). For each accepted pair \((z,\phi)\), the two-body \(K^0\) and \(\Lambda\) four-vectors were constructed in the center-of-mass frame and boosted to the laboratory frame.

Every accepted production event was forced through both visible channels $K_S^0\to\pi^+\pi^-$ and $\Lambda\to p\pi^-$. The daughter particles were generated isotropically in the corresponding parent rest frames using standard phase-space routines~\cite{Cowan:2010js}, and were subsequently transformed to the laboratory frame. For fixed-mass two-body decays the phase-space weight is constant, so this procedure generates an isotropic decay without an additional event weight. Because the present production model contains only one real vector amplitude and predicts \(P_y=0\), the \(\Lambda\) decay was also taken to be isotropic in its rest frame.

The branching fractions in Eq.~\eqref{eq:visible-branching-fraction} were not sampled in the event generation. Since the \(5 \times 10^7\) entries represent a sample conditional on the selected \(K_S^0\to\pi^+\pi^-\) and \(\Lambda\to p\pi^-\) channels, they cannot be interpreted as an absolute yield. In addition, the simulation is sufficient to used to investigate the kinematics, even though the neutral-kaon mixing, particle lifetimes, and displaced decay vertices were likewise not simulated.

This level of description is sufficient to reproduce the truth-level momentum--angle plots. A quantitative feasibility study should
add a detector-transport simulation, trigger and reconstruction algorithms, background channels and so on. For comparison,
the J-PARC E40 analysis generated the production and decay chain with a realistic \(\Lambda\) branching fraction and used a Geant4-based simulation to evaluate acceptance, tracking, and energy-measurement efficiencies~\cite{J-PARCE40:2025hhw}.

\subsection{Laboratory momentum--angle distributions}

\begin{figure*}[htbp]
  \centering
  \includegraphics[scale=0.8]{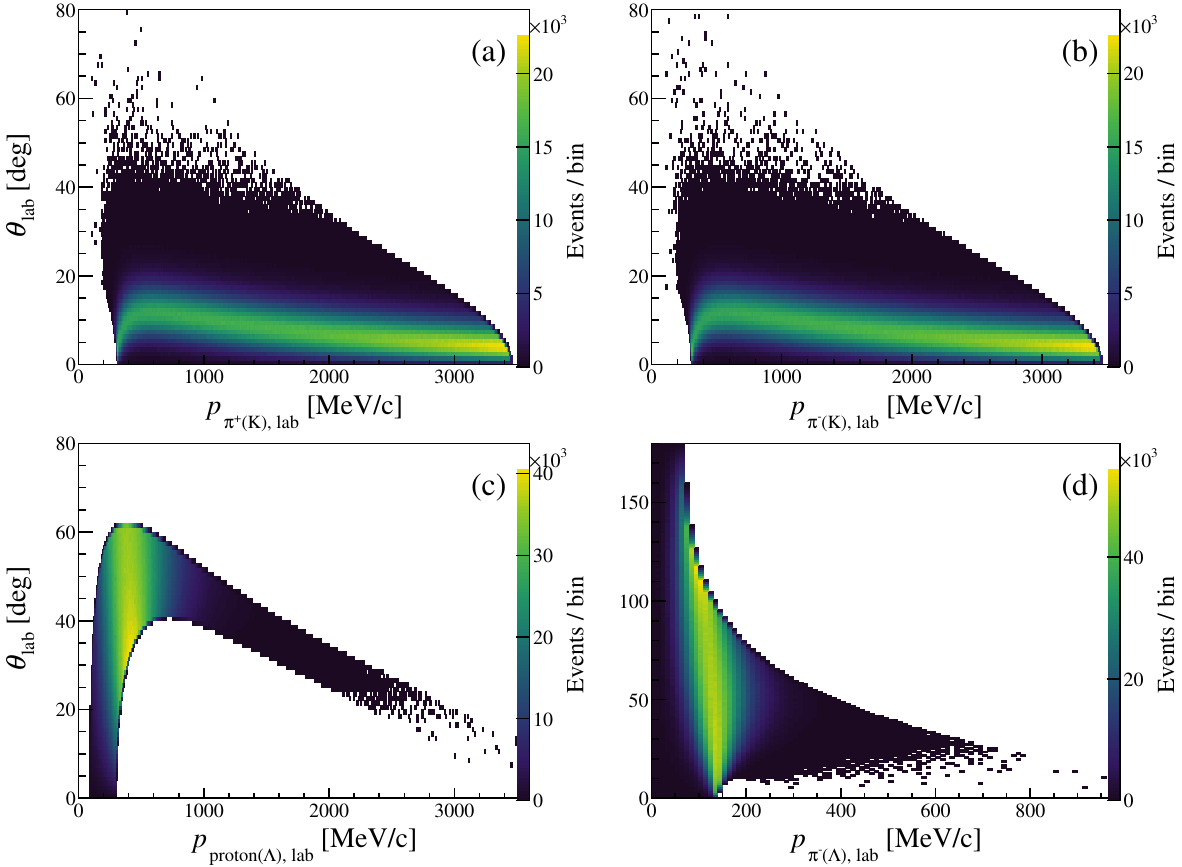}
  \caption{Idealized laboratory momentum versus polar angle for the charged daughter particles in a conditional sample of \(5\times10^7\) generated \(\reaction\) events at \(p_{\pi,\lab}=4~\gev/c\). Panels (a) and (b) show the \(\pi^+\) and \(\pi^-\) from \(K_S^0\to\pi^+\pi^-\), respectively, while panels (c) and (d) show the proton and \(\pi^-\) from \(\Lambda\to p\pi^-\), respectively. The bin widths are \(10~{\rm MeV}/c\) in momentum and \(1~\degree\) in polar angle. 
  }
  \label{fig:cascade-analysis3}
\end{figure*}

Using the event generation algorithm described in Sec.~\ref{subsec:3B}, $5 \times 10^7$ events were generated. It's therefore significant to inspect the kinematic distributions of the final-state particles that would be measured directly by the detectors. Fig.~\ref{fig:cascade-analysis3} shows the laboratory momentum and polar angle of the four charged particles with \(p_{\pi,\lab}=4~\gev/c\) beam. The two pions from the neutral kaon are concentrated at small angle. Their mean momenta are about \(1.8~\gev/c\), and their mean polar angles are about \(8.4^\circ\). This forward collimation follows from the strong forward production of the \(K^0\) and the Lorentz boost of an isotropic two-body decay.

The proton from the \(\Lambda\) decay has a different pattern. Its distribution forms a broad arch in the \((p_{\lab},\theta_{\lab})\) plane,
with a mean momentum of about \(0.4~\gev/c\) and a mean polar angle of about \(43.5^\circ\). This shape is a kinematic boundary rather than a resonance or detector effect. In the \(\Lambda\) rest frame, the proton momentum is fixed:
\begin{equation}
  p_p^*=\frac{\lambda^{1/2}
  (m_\Lambda^2,m_p^2,m_\pi^2)}{2m_\Lambda}
  \simeq100.6~\mev/c.
  \label{eq:lambda-decay-momentum}
\end{equation}
Regarding a decay emitted at rest-frame angle \(\theta^*\), as \(\cos\theta^*\) varies from \(-1\) to 1, the same fixed rest-frame
momentum is mapped into correlated laboratory momenta and angles with the specific velocity of \(\Lambda\) along its local longitudinal 
direction. The spread of the produced \(\Lambda\) momentum and direction broadens this one-parameter kinematic curve into the observed arch-shaped band. The scarcity of high-momentum protons at very small polar angles is a consequence of the production direction and two consecutive boosts.

The \(\pi^-\) from the \(\Lambda\) decay is much softer. Although it has the same rest-frame momentum magnitude as the proton, its rest-frame energy is much smaller, so the boost produces a broad angular distribution with a mean momentum of about \(124.5~\mev/c\) and a mean polar angle near \(68.0^\circ\). This particle is likely to set the demanding low-momentum and large-angle tracking requirement. In contrast, the forward \(K_S^0\) daughters require coverage close to the beam direction over a broad momentum interval extending to approximately \(3.4~\gev/c\). A future detector concept should therefore combine forward high-momentum tracking with efficient reconstruction of soft, large-angle pions and intermediate-angle protons. 

The simulation also clarifies the experimental strategy for testing the production model. Acceptance must be evaluated in at least the variables \((p,\theta)\) for every daughter particles, because a one-dimensional efficiency can distort the strongly correlated forward distribution. After reconstruction, the kaon center-of-mass angle, $\cos{\theta^{\cm}_K}$, and Mandelstam variable \(t\) should be unfolded in several beam momentum bins and compared directly with the differential cross section distribution in Fig.~\ref{fig:dsigma}. In addition, the total cross section and momentum-angle distributions of final-state particles can be compared with the corresponding measurements, as shown in Fig.~\ref{fig:total-fit} and Fig.~\ref{fig:cascade-analysis3}. These comparisons would give a complementary test for the production model proposed in this work.

To estimate the kinematic acceptance of \(\Lambda\to p\pi^-\) decays at HHaS, we examine the generated events individually. The unpolarized sample is used as a baseline, assuming that polarization has a negligible effect on the integrated acceptance. Following the nominal charged-particle acceptance of HHaS~\cite{Chen:2025ppt}, both the proton and the pion are required to satisfy
\begin{equation}
  10^\circ < \theta_{\rm lab} < 100^\circ,
  \qquad p_T > p_T^{\rm min},
\end{equation}
with full azimuthal coverage, where $\theta_{\rm lab}$ represents the polar angle of the emitted particles in the laboratory frame. Here,
the transverse momentum \(p_T^{\rm min}=50~\mathrm{MeV}/c\) for the magnetic field \(B=1.5~\mathrm{T}\) and \(p_T^{\rm min}=130~\mathrm{MeV}/c\) for \(B=4~\mathrm{T}\). The acceptance is defined as
\begin{equation}
  A_\Lambda =
  \frac{N_\Lambda^{\rm pass}}{N_\Lambda^{\rm gen}},
\end{equation}
where \(N_\Lambda^{\rm gen}\) is the total number of generated \(\Lambda\) hyperons and \(N_\Lambda^{\rm pass}\) counts those whose proton and pion both satisfy the acceptance requirements. All generated \(\Lambda\) hyperons are forced to decay through the \(p\pi^-\) channel, so the branching fraction is not included in this ratio. No acceptance requirement is imposed on the accompanying \(K^0\) or its decay products.
At \(p_{\pi,\rm lab}=4~\mathrm{GeV}/c\), the resulting acceptance is \(77.4\%\) for \(B=1.5~\mathrm{T}\) and \(36.3\%\) for \(B=4~\mathrm{T}\). These fractions quantify the kinematic availability of \(\Lambda\) decays and further polarization studies.

\section{Summary}  \label{sec:summary}

We studied \(\reaction\) under the working hypothesis that Reggeized \(t\)-channel \(K^*(892)\) exchange, which gives the leading contribution at incident pion momenta from 2.6 to \(6.0~\gev/c\). The charged \(K^*K\pi\) coupling was fixed to \(-4.27\), the vector
\(K^*\Lambda N\) coupling to \(-6.19\) while the tensor coupling was omitted. With only the cutoff \(\Lambda_{\rm cut}\) varied, 13
total-cross-section points were described with \(\Lambda_{\rm cut}=1381.2 \pm 30.7~\mev\) and \(\chi^2/N_{\mathrm{dof}}=1.0\). For comparison, we also consider the conventional Feynman propagator for \(K^*(892)\) exchange. The results show that the Reggeized propagator provides a better description of the experimental data of total cross section.

Measurements of differential cross section at several beam momenta are required to test both the predicted forward slope and the Regge energy dependence. A nonzero normal \(\Lambda\) polarization would provide a separate indication of an omitted relative phase, although no polarization information was used here since it's limited impact on the kinematics of final-state particles statistically.

Moreover, we try to give some implications following our theoretical results. Using free-proton cross section and the proton areal densities of the targets, as well as the incident pion with \(p_{\pi, {\rm lab}} = 4~\gev/c\), we estimate elementary production probabilities of
\(3.13\times10^{-5}\) and \(2.62\times10^{-5}\) for 10-cm liquid-deuterium and liquid-hydrogen targets, respectively. The final reconstructed yield remains a function of the integrated pion flux, branching fractions, neutral-kaon propagation, decay-volume acceptance, and detector efficiency. Applying \(10^8 {\rm s^{-1}}\) incident pion beam, the ideal visible rates before these acceptance and efficiency losses are about \(6.89\times10^2\) and \(5.76\times10^2\) events/s, respectively. The idealized simulation shows that the two \(K_S^0\) decay pions are strongly forward, the \(\Lambda\)-decay proton occupies a broad momentum--angle arch, and the \(\Lambda\)-decay pion is soft and extends to large angles. These complementary kinematic regions provide concrete requirements for the acceptance and tracking studies needed before a quantitative experimental proposal can be made.

\section*{Acknowledgement}

This work is partly supported by the National Key R\&D Program of China under Grant No. 2023YFA1606703 and the National Natural Science Foundation of China under Grant Nos. 12435007, 12361141819, and 12575094.

\normalem
\bibliography{ref}

\end{document}